\documentstyle[12pt,epsf,amssymb]{article}

\begin{document}

\baselineskip=18.8pt plus 0.2pt minus 0.1pt

\def\CR{\nonumber \\}
\def\pt{\partial}
\def\be{\begin{equation}}
\def\ee{\end{equation}}
\def\bea{\begin{eqnarray}}
\def\eea{\end{eqnarray}}
\def\eq#1{(\ref{#1})}
\def\la{\langle}
\def\ra{\rangle}
\def\hyp{\hbox{-}}

\begin{titlepage}
\title{
\hfill\parbox{4cm}
{\normalsize YITP-02-2 \\{\tt hep-th/0201130}}\\
\vspace{1cm}
A de-Sitter thick domain wall solution \\ by elliptic functions}
\author{
Naoki {\sc Sasakura}\thanks{\tt sasakura@yukawa.kyoto-u.ac.jp}
\\[7pt]
{\it Yukawa Institute for Theoretical Physics, Kyoto University,}\\
{\it Kyoto 606-8502, Japan}}
\date{\normalsize January, 2002}
\maketitle
\thispagestyle{empty}

\begin{abstract}
\normalsize
We obtain and study an analytical solution of a de-Sitter thick
domain wall in five-dimensional Einstein gravity interacting with a scalar field. 
The scalar field potential is axion-like, 
$V(\phi)=a+b \cos(\sqrt{2/3}\phi)$ with constants $a,b$ 
satisfying $-3b<5a<3b$, 
and the solution is expressed in terms of elliptic functions. 
\end{abstract}
\end{titlepage}

Solitons play major roles in non-perturbative aspects of 
field theory and string theory.
Since string theory contains gravity, solitons necessarily couple with gravity
in string theory.
Because of the non-linearity and instability of the gravitational interactions,
the inclusion of the gravitational dynamics into a soliton dynamics 
is a highly non-trivial problem.
The simplest soliton interacting with gravity would be a domain wall in
a coupled system of a scalar field and gravity. 
The easiest to treat are the BPS domain walls in supergravity theories, 
in which the equations of motions are given by the BPS first order 
differential equations \cite{Cvetic}. These solutions are static. In fact,
it turns out that static domain walls without supersymmetry can also 
be treated in the same manner \cite{Csaki,Skenderis,Chamblin,DeWolfe}.
A solution of a domain wall with a de-Sitter expansion was 
first constructed in the thin wall approximation \cite{Vilenkin:1983hy}.
Not so many analytic solutions of a thick domain wall with a de-Sitter
expansion are known \cite{Gremm:2000dj,Wang:2002pk}.
In this paper, we obtain another analytic solution of a thick domain wall
with a de-Sitter expansion.    

The system we discuss in this paper is the coupled system of gravity 
and a scalar field in five-dimensions, and the action is given by 
\be
S=\int dt d^3x dy 
\sqrt{-g} \left(R - \frac12 g^{\mu\nu} \pt_\mu \phi \pt_\nu \phi
-V(\phi)\right).
\ee
The metric ansatz we use is the warped geometry
\be
\label{warplor}
ds^2=a(y)^2 (-dt^2+e^{2Ht}(dx^i)^2)+dy^2,
\ee
where $H$ denotes the Hubble constant of the four-dimensional de-Sitter
space-time.
Under the assumption that the scalar field depends only on the coordinate $y$,
the Einstein equations are
\bea
\label{eeq1}
V(\phi)&=&\frac{-3 a a''-9 (a')^2+9 H^2}{a^2}, \cr
(\phi')^2&=& \frac{6(a')^2-6 a a''-6H^2}{a^2},
\eea
where $'$ denotes the derivation with respect to $y$.
The equation of motion of the scalar field is automatically satisfied by 
the solutions of \eq{eeq1} because of Bianchi identity.

For $H>0$, the second equation of \eq{eeq1} gives an inequality 
\be
\label{ineq}
(a')^2-a a''\ge H^2 > 0.
\ee
This inequality constrains the possible shape of a domain wall 
solution. At the peak of a domain wall defined by $a'=0$, this inequality
imposes $a''<0$. Thus there can exist at most one peak, and $a$ is a 
monotonically decreasing function of the distance from the peak.
If we assumed that $a$ was bounded from below, $a'$ and $a''$ would approach
zero in the limit $|y|\rightarrow \infty$, and would violate the inequality. 
Thus $a$ must cross $a=0$ at finite $y$'s on the both sides of a domain wall.
The behavior of $a$ near the vanishing point has a physical significance.
Without tuning the behavior of $a$, 
\eq{eeq1} shows that the sources of the energy-momentum $V(\phi)$ and $\phi'$
diverge at the vanishing point and the point becomes a naked singularity.
To avoid this situation \footnote{However see \cite{Gremm:2000dj}, 
where it is argued that solutions with singularities may also be 
physically meaningful in view of AdS/CFT correspondence.}, 
the behavior of $a$ near the vanishing point is constrained to have the form
\be
\label{behavea}
a(y)=-Hy+O(y^3),
\ee
where we are considering a domain wall whose peak is located 
at a negative value of $y$ and that a vanishing point is located at $y=0$.
Note that this behavior of $a$ is similar to that of a Rindler horizon.
In a series of the solutions presented in \cite{Wang:2002pk}, the vanishing
points are horizons and the extended space-times are obtained.      
In this paper we will obtain and study another analytic solution 
of a de Sitter domain wall which is regular in the sense that 
the vanishing points have the behavior \eq{behavea}.

As discussed in \cite{DeWolfe}, the second order differential equations 
\eq{eeq1} can be 
rewritten in the following BPS-like first order differential equations 
by introducing a ``superpotential'' $W(\phi)$:
\bea
\label{firsteqoth}
\phi'&=&\frac3\gamma \frac{\pt W(\phi)}{\pt \phi},\cr
\frac{a'}{a}&=&-\frac12\gamma W(\phi), \cr
\gamma&=&\sqrt{1+\frac{4 H^2}{a^2 W^2}},
\eea
and the scalar field potential is given by 
\be
\label{firsteq}
V(\phi)=-3 W(\phi)^2+\frac9{2 \gamma^2} 
\left( \frac{\pt W(\phi)}{\pt \phi}\right)^2.
\ee
Although this form of the differential equations is very convenient
in the study of some general aspects of solutions and will be used later, 
another form discussed in \cite{Flanagan:2001dy} is more convenient 
for the purpose of obtaining explicit solutions.
With the substitution $a=\exp(-A/2)$, the Einstein equations become
\bea
\label{anotherform}
V&=&-3 A'^2+\frac32 A''+9 H^2 \exp(A), \cr
\phi'^2&=&3 A''-6 H^2 \exp(A).
\eea
A solution of these differential equations is a well-defined function of $y$. 
By solving $y$ as a function of $A$ locally, 
one may regard the solution as a function of $A$ rather than $y$ with some
care on multi-value.    
Now let us define $g(A)$ by
\be
\label{defineg}
\phi'^2=\frac32 \frac{dg(A)}{dA},
\ee
and integrate the second equation of \eq{anotherform} after multiplying $A'$.
Then we obtain 
\be
\label{solg}
g(A)=A'^2-4 H^2 \exp(A).
\ee
Here the constant shift ambiguity of the definition \eq{defineg} of $g(A)$ 
has been used to eliminate the integration constant in \eq{solg}.  
The $g(A)$ is related to the previous form of the differential equations by
\be
\label{gandw}
g(A)=W(\phi)^2.
\ee
Thus the set of differential equations we will solve is given by
\bea
\label{diffeqs}
\phi'^2&=&\frac32 \frac{dg(A)}{dA},\cr
A'^2&=& g(A)+4 H^2 \exp(A), \cr
V&=&-3 g(A)+\frac34 \frac{dg(A)}{dA}.
\eea
  
For each choice of $g(A)$ we obtain a solution \cite{Flanagan:2001dy}. 
We take the choice
\be
\label{formg}
g(A)=-\frac{4 H^2}{\beta^2}(a^2+\alpha),
\ee
where $\alpha$ and $\beta$ are real constants and $a=\exp(-A/2)$ as above.
The first two equations of \eq{diffeqs} give
\bea
\label{diffeqforaphi}
\frac{d\phi}{d Y}&=&\sqrt{6} a, \cr
\frac{da}{d Y}&=&-\sqrt{-a^4-\alpha a^2+\beta^2},
\eea
where $Y=\frac{H}\beta y$, and we have chosen the signatures of 
the square roots corresponding to 
the right-hand side (larger $y$) region of the domain wall 
where $\phi$ increases and $a$ decreases.
The all over factor of $a$ determines the length unit and we may fix it
by setting $a=1$ at the peak of the domain wall defined by $a'=0$.
Hence we take $\alpha=\beta^2-1$. 
The solutions to \eq{diffeqforaphi} are expressed by elliptic functions. 
Explicitly, we obtain
\bea
\label{solution}
a(y)&=&-{\rm sn}(H y,i\beta^{-1}),\cr
\phi(y)&=&\sqrt{6} {\rm Arctan}
\left(\frac{{\rm cn}(H y,i\beta^{-1})}{\beta {\rm dn}(H y,i\beta^{-1})}\right),
\eea
where we have fixed the integration constants so that 
a vanishing point $a=0$ is located at $y=0$ 
and the scalar field takes $\phi=0$ at the peak of the domain wall. 
The elliptic functions are defined
by the inverse of the elliptic integral
\be
{\rm sn}^{-1}(z,k)=\int_0^z\frac{dx}{\sqrt{(1-x^2)(1-k^2x^2)}},
\ee   
and ${\rm cn}(u,k)=(1-{\rm sn}^2(u,k))^{1/2}$,
${\rm dn}(u,k)=(1-k^2{\rm sn}^2(u,k))^{1/2}$.
By solving $a(\phi)$ 
and substituting this into the third equation of \eq{diffeqs}, the potential
is given by 
\be
\label{potential}
V(\phi)=\frac{9H^2(\beta^2-1)}{2\beta^2}+\frac{15H^2(\beta^2+1)}{2 \beta^2}
\cos\left(\sqrt{\frac23}\phi\right).
\ee
This scalar potential has the form of that of an axion 
with an instanton correction. 
 
By taking certain limits, our solution agrees with some of the known solutions.
For $H=c \beta\rightarrow 0$, the solution represents a static domain wall 
with $H=0$ and $W(\phi)=2 c \sin (\phi/\sqrt6)$ in \eq{firsteqoth} and 
\eq{firsteq}.
This ``superpotential'' was recently derived for a hypermultiplet in 
${\cal N}=2$ gauged supergravity derived from a non-homogeneous quaternionic
space \cite{Behrndt:2001km}.
This interesting development suggests a possible concrete route for the 
inclusion of our model into string- or M-theory, in spite of the widely 
applicable no-go theorem of \cite{Maldacena:2001mw}.  
For $\beta=1$, the solution agrees with one of the solutions presented
in \cite{Wang:2002pk}.
This can be checked by changing to the conformal coordinate $dz=dy/a$.
For $\beta\rightarrow \infty$, the scalar field freezes at $\phi=0$ 
and the solution is just for a five-dimensional cosmological 
constant \cite{Kaloper:1999sm}.
An analytic continuation to $\beta^2<0$ does not work because $\phi(y)$ is ill defined. 

It is interesting to study how a static domain wall 
responds to an additional energy-momentum source. 
This kind of analysis was done for an energy-momentum source localized 
on a thin wall, and the standard cosmological evolution was 
reproduced \cite{Csaki:1999jh,Cline:1999ts} 
in the Randall-Sundrum approach \cite{Randall:1999vf}.
We can give a qualitative discussion on this question for a thick domain wall
by interpreting judiciously our solution.
Let us consider the change of the constant part of the potential 
\eq{potential} from that of the static case with 
$W(\phi)=2 c \sin (\phi/\sqrt6)$:
\be
V(\phi)=-\frac{9c^2}{2}+\delta \Lambda+\frac{15 c^2}{2} 
\cos\left(\sqrt{\frac23}\phi\right),
\ee
where $\delta \Lambda$ is a positive constant. 
Comparing with \eq{potential}, we have
\bea
\delta \Lambda &=&  \frac{9H^2(\beta^2-1)}{2\beta^2}+\frac{9c^2}{2}, \cr
0&=&\frac{15H^2(\beta^2+1)}{2 \beta^2}-\frac{15 c^2}{2}.
\eea
Solving the above equations we obtain
\be
\label{handlambda}
H^2=\frac{\delta \Lambda}9.
\ee
Because of the redshift factor, the additional energy-momentum generated by
$\delta \Lambda$ is well concentrated near the wall, 
and hence $\delta \Lambda$ may be regarded as a source of the 
cosmological constant on the wall. 
Hence the relation \eq{handlambda} may be interpreted as $H\sim \sqrt{\rho}$,
which seems to be a good sign for the standard cosmological evolution  
for a thick domain wall.

In the above analysis of obtaining \eq{handlambda}, the
regularity constraint \eq{behavea} is essential.
If we allowed a naked singularity to appear, 
the expansion rate $H$ could not be related to $\delta \Lambda$.
In fact we can construct a solution with $H=0$ for any $\delta \Lambda$.
For $\gamma=1$ ($H=0$),
\eq{firsteq} is a first order differential equation for $W(\phi)$:
\be
\label{diffeqforw}
\frac{dW(\phi)}{d\phi}=\sqrt{\frac29 V(\phi)+\frac23 W(\phi)^2}.
\ee
Since $a'=0$ at the peak of the wall, the trajectory of the solution 
goes through $(\phi,W)=(0,0)$ in the $(\phi,W)$ plane.
The solution $W(\phi)=2 c \sin (\phi/\sqrt6)$ for the static case 
$\delta \Lambda=0$ is a fine-tuned solution, and one can see that 
a slight change of the potential $V(\phi)$ will 
result in a divergence of the asymptotic behavior of $W(\phi)$ in large 
$\phi$ as observed in \cite{DeWolfe}. 
For most cases including the above modified potential, 
the trajectory of the solution in large $\phi$ enters the region where 
the potential energy $V(\phi)$ is negligible in \eq{diffeqforw}. Thus
\be
\label{asymw}
W(\phi)\sim \exp\left(\sqrt\frac23\phi\right)
\ee
for large $\phi$. This region is where the scalar kinetic energy dominates
over the potential energy.
From \eq{asymw} and \eq{firsteqoth}, we have
\be
a(y)\sim (-y)^{1/4},
\ee
and this is a naked curvature singularity.
Thus in our solution \eq{solution}, the curvature singularities are 
regularized by a de-Sitter expansion. 
This mechanism to use a de-Sitter expansion to turn a curvature singularity
of a static soliton to a horizon has also appeared in the context of global
$U(1)$ vortex solutions \cite{Gregory:1996dd} and 
in a codimension two non-supersymmetric soliton solution in IIB string 
theory \cite{Berglund:2001aj}.
It would be very interesting to extend the analysis of the response of 
a thick domain wall to a more general energy-momentum source
under the regularity constraint \eq{behavea}. 

The extension to the other cases than \eq{formg} would also be interesting.
To avoid a curvature singularity at $a=0$, the energy-momentum 
sources $V$ and $\phi'^2$ should not diverge in the limit $a\rightarrow 0$.
This is equivalent to the condition that both $g(A)$ and $dg(A)/dA$ 
do not diverge in the limit $A\rightarrow\infty$.
Thus a general choice $g=c_1a^2+c_2 a+c_3$ is also allowed for a regular 
solution, and we will obtain an explicit expression in terms of
elliptic functions.
In this case a non-trivial issue is whether the scalar field potential 
has an analytic expression in terms of $\phi$ or not.

Finally we will study the linear perturbations around our solution and
prove its stability.
The problem of obtaining the mass spectrum boils down to solving a 
Schrodinger equation 
\be
\label{scheq}
\left(-\frac{d^2}{dz^2}+V_{t,e}(z)\right)\varphi = m^2 \varphi,
\ee
where $V_{t}$ and $V_{e}$ are the potential energies 
for the tensor perturbation \cite{DeWolfe,Gremm:2000dj,Kobayashi:2001jd}, 
\bea
ds^2&=&a^2((\gamma_{\mu\nu}+h_{\mu\nu})dx^\mu dx^\nu+dz^2),\cr
{h_{\mu}}^\mu&=&{h_{\mu\nu}}^{|\nu}=0,
\eea
and the scalar perturbation,
\be
ds^2=a^2((1+\psi_1)\gamma_{\mu\nu}dx^\mu dx^\nu+(1+\psi_2)dz^2),
\ee
respectively \cite{Kobayashi:2001jd,DeWolfe:2000xi}. 
As for the tensor perturbation, the potential is given by 
\be
\label{tensor}
V_t=\frac32 \frac{d{\cal H}}{dz}+\frac94 {\cal H}^2,
\ee
where
\be
{\cal H}=\frac1a\frac{da}{dz}.
\ee 
The Schrodinger equation \eq{scheq} can be rewritten in the 
form \cite{DeWolfe}
\be
\label{qqeq}
Q^\dagger Q\varphi=m^2\varphi,
\ee
where 
\be
Q=\frac{d}{dz}-\frac32 {\cal H}.
\ee
Thus there is a normalizable state $\varphi=a^{3/2}$ with $m^2=0$,
and the other states have $m^2>0$.
As discussed generally in \cite{Gremm:2000dj},
our solution is stable against the tensor perturbation.
Substituting our solution \eq{solution} into \eq{tensor},
we obtain
\be
V_t=
\frac{3 H^2}{4 \beta^2} \left(3 \beta^2+5(1-\beta^2)a^2-7 a^4 \right).
\ee
This potential has a well near the domain wall peak and the low-energy states 
trapped in the well play the role as the graviton on the brane world.

The potential for the scalar perturbation has the 
expression \cite{Kobayashi:2001jd}
\be
\label{vepot}
V_e=-\frac52 {\cal H}^{(1)}+\frac94 {\cal H}^2 +{\cal H} 
\frac{\phi^{(2)}}{\phi^{(1)}}-\frac{\phi^{(3)}}{\phi^{(1)}}
+2\left(\frac{\phi^{(2)}}{\phi^{(1)}}\right)^2 - 6H^2,
\ee
where $(i)$ denotes the $i$-th derivative with respect to $z$.
Using \eq{firsteqoth}, this potential can be expressed by $a$, $W$ 
and $\gamma$. Assuming a form 
${\cal I}=d_1 a \gamma W+d_2 (a/\gamma) \pt^2_\phi W+d_3 a \gamma'/\gamma$
and after a straightforward tedious computation,  
we arrive at another expression 
\be
\label{ve}
V_e=-\frac{d{\cal I}}{dz}+{\cal I}^2+\frac{a^2(\phi')^2}3-4 H^2,
\ee
where ${\cal I}$ turns out to be 
\be
{\cal I}=\frac {d \ln (a^{3/2}\phi')}{dz}.
\ee
The expression \eq{ve} shows that the Schrodinger equation for the 
scalar perturbation does not seem to have a non-negative expression 
like \eq{qqeq}.
For $H=0$ it is evident that the remainder is non-negative, 
but for $H>0$ we cannot expect this in general. 
Especially, it is not so for our solution.   
Hence the question of stability whether there exist tachyonic states or not 
seems to need a case-by-case study.   
Substituting \eq{solution} into the expression \eq{ve}, we obtain
\be
V_e=\frac{3 H^2}{4 \beta^2}\left( 3 \beta^2 +5(1-\beta^2) a^2+a^4\right).
\ee
In the example studied in \cite{Kobayashi:2001jd}, 
the potential $V_e$ is positive in the whole range of $z$ 
and takes the maximum value at the domain wall peak.
Thus it is evident that there does not exist any tachyonic
or $m=0$ normalizable states in their example.
In our example, however, the situation is more complicated. For $\beta^2>3$, 
there is a potential well with negative values near the domain wall peak.
Hence it seems hard to answer the question of stability only from the 
shape of the potential.
But fortunately, we have a useful relation
\be
V_e=V_t+\frac{6 H^2 a^4}{\beta^2}.
\ee
Since we know that $-d^2/dz^2+V_t$ is non-negative,  
$-d^2/dz^2+V_e$ is positive definite.
Thus our solution is stable against the scalar perturbation.

\vspace{.5cm}
\noindent
{\large\bf Acknowledgments}\\[.2cm]
The author would like to thank K.~Behrndt and G.~Dall'Agata for
valuable comments on \cite{Behrndt:2001km}. 
The author was supported in part by Grant-in-Aid for Scientific Research
(\#12740150), and in part by Priority Area:
``Supersymmetry and Unified Theory of Elementary Particles'' (\#707),
from Ministry of Education, Science, Sports and Culture, Japan.

\end{document}